\documentstyle[12pt,epsf]{article}

\def\beeq{\begin{eqnarray}} \def\eeeq{\end{eqnarray}}
\newcommand\mysection{\setcounter{equation}{0}\section}
\renewcommand{\theequation}{\thesection.\arabic{equation}}
\newcounter{hran} \renewcommand{\thehran}{\thesection.\arabic{hran}}

\def\bmini{\setcounter{hran}{\value{equation}}
  \refstepcounter{hran}\setcounter{equation}{0}
  \renewcommand{\theequation}{\thehran\alph{equation}}\begin{eqnarray}}

\def\bminiG#1{\setcounter{hran}{\value{equation}}
\refstepcounter{hran}\setcounter{equation}{-1}
\renewcommand{\theequation}{\thehran\alph{equation}}
\refstepcounter{equation}\label{#1}\begin{eqnarray}}

\def\emini{\end{eqnarray}\relax\setcounter{equation}{\value{hran}}\renewcommand{\theequation}{\thesection.\arabic{equation}}}

\def\ben{\begin{enumerate}}  \def\een{\end{enumerate}}
\def\bit{\begin{itemize}}    \def\eit{\end{itemize}}
\def\beq{\begin{equation}}   \def\eeq{\end{equation}}
\def\bea{\begin{eqnarray}}  \def\eea{\end{eqnarray}}
\def\nn{\nonumber}
\def\noi{\noindent}

\def\lsim{\raise0.3ex\hbox{$<$\kern-0.75em\raise-1.1ex\hbox{$\sim$}}}
\def\gsim{\raise0.3ex\hbox{$>$\kern-0.75em\raise-1.1ex\hbox{$\sim$}}}
 \def\cite#1{[\ref{#1}]}
 \def\citd#1#2{[\ref{#1},\ref{#2}]}
 \def\citt#1#2#3{[\ref{#1},\ref{#2},\ref{#3}]}
 \def\citm#1#2{[\ref{#1}--\ref{#2}]}
 \def\citr#1#2#3#4{[\ref{#1},\ref{#2},\ref{#3}--\ref{#4}]}

\begin{document}
\begin{center}
{\bf FIELD STRENGTH CORRELATOR AND AN INFRARED FIXED POINT OF
THE WILSONIAN EXACT RENORMALIZATION GROUP EQUATIONS} \\
\vspace{1 truecm}
{\bf Ulrich Ellwanger}\footnote{email : ellwange@qcd.th.u-psud.fr}\\
Laboratoire de Physique Th\'eorique et Hautes
Energies\footnote{Laboratoire associ\'e au Centre National de la
Recherche Scientifique 
(URA D0063)}\\    Universit\'e de Paris XI, Centre d'Orsay, B\^atiment
210, 91405 
Orsay Cedex, France\\  
\end{center}
\vspace{3 truecm}
\begin{abstract}
\noindent
The correlator of two field strengths is computed from an effective
action for Yang Mills theories, which contains both gluons and an auxiliary
antisymmetric tensor field for the field strength as local
variables. This action allows to relate explicitly many different
approaches to confinement, and is computed using Wilsonian
renormalization group equations with the bare Yang Mills action as 
starting point. Due to the inclusion of a higher dimensional operator in
the ghost sector the running gauge coupling becomes vanishingly small at
a critical scale $k_c$, and the resulting low energy action resembles
the action of a confining string theory proposed by Polyakov.  
\end{abstract}

\vspace{2 truecm} 

\noi LPTHE Orsay 98-48 \\ 
\noi July 1998 \\
 
\newpage
\pagestyle{plain}
\mysection{Introduction}
\hspace*{\parindent}
Many different methods have been used in order to characterize the
infrared behaviour of Yang Mills theories: the Wilson loop \cite{1r},
the infrared behaviour of the gluon propagator \cite{2r}, the monopole
condensate from a dual Higgs theory \cite{3r} and corresponding analytic
models \cite{4r}, a confining string theory \cite{5r} and field 
strength correlators \cite{6r}. The area law of the Wilson loop has been
related to the infrared behaviour of the gluon propagator in \cite{7r},
to monopole condensates in the dual Higgs theory in \citd{3r}{4r} and to
field strength correlators in \cite{6r}. More recently, field strength
correlators have been studied in the dual Higgs theory in \cite{8r} and
in the confining string theory in \cite{9r}. \par 

In the present paper we continue the study of an effective action for
Yang Mills theories, which contains both gluons and an auxiliary
antisymmetric tensor field for the field strength as local variables
\cite{10r}. It has the virtue of allowing for an explicit duality
transformation of its abelian projection, and allows thus to relate 
explicitly a $1/q^4$ behaviour of the gluon propagator to a monopole
condensate in a dual Higgs theory. Moreover, the parameters of this
effective action can be computed from the bare Yang Mills action by
integrating the Wilsonian exact renormalization group equations. Note
that antisymmetric tensor fields have already appeared frequently in
effective descriptions of the infrared behaviour of Yang Mills theory,
motivated, e.g., by confining string theories \cite{5r} and, again,
duality transformations of dual Higgs theories. \par

The aim of the present paper is twofold: first, in chapter 2, we compute
the correlator of two field strengths from the corresponding effective
action. We find indeed an exponential decrease at large distances, in
agreement with results on the lattice \cite{11r}, from the dual Higgs
model \cite{8r} and from confining strings \cite{9r}. Notably we find
that the parameters which characterize the slope of the exponential
decrease are not the ones which parametrize a $1/q^4$ behaviour of the 
gluon propagator. \par

Second, we reconsider the Wilsonian renormalization group flow of the
parameters of the Yang Mills action in the infrared regime in chapter
3. All previous approaches in this direction \citd{12r}{13r} have been
plagued with the appearance of a Landau singularity of the running
coupling constant, which did not allow the integration of the
renormalization group equations with respect to an infrared cutoff $k^2$
down to $k^2 = 0$ (unless the running coupling is put to a finite value
at $k^2 = 0$ by hand). By taking a higher dimensional operator in the
effective action (in the ghost sector) into account, we find that the
Landau singularity is avoided. 
Moreover, we obtain an effective action in the deep infrared regime, which
corresponds to the one of an effective abelian theory and resembles the
action of a confining string theory \cite{5r}. Conclusions and a
discussion of the physical interpretation of this result will be given
in chapter 4.     

\mysection{Effective Action and the Field Strength Correlator}
\hspace*{\parindent}
In order to define the effective action with an auxiliary field for the
field strength we start with the Yang Mills partition function 

\beq
e^{-G(J, \chi , \bar{\chi})} = {1 \over N} \int {\cal D} (A, c , \bar{c}) \
e^{-S_{YM} - S_g + J \cdot A + \bar{\chi} \cdot c + \chi \cdot \bar{c}}
\quad .  
\label{2.1} 
\eeq

\noi Here $S_{YM}$ is the standard Yang Mills action,

\beq
S_{YM} = {1 \over 4} \int d^4x \ F_{\mu \nu}^a \ F_{\mu \nu}^a \quad , 
\label{2.2}
\eeq

\noi and $S_g$ is the gauge fixing and ghost part:

\beq
\label{2.3}
S_g = \int d^4x \left [ {1 \over 2 \alpha} \left ( \partial_{\mu}
A_{\mu}^a \right )^2 + \partial_{\mu} \bar{c}^a \left ( \delta_{ac} \
\partial_{\mu} + g \ f_{abc} \ A_{\mu}^b \right ) c^c \right ] \quad
. \eeq 

\noi Next we multiply the right-hand side of eq. (\ref{2.1}) with

\beq
1 = {1 \over N'} \int {\cal D} H \ e^{-{1 \over 4} \int d^4x(F_{\mu
\nu}^a - H_{\mu \nu}^a)^2} \quad . \label{2.4}
\eeq 

\noi In addition we add a source $K_{\mu \nu}^a$ for the auxiliary field
$H_{\mu \nu}^a$ and obtain

\beq
\label{2.5}
e^{-G(J, K, \chi , \bar{\chi})} = {1 \over NN'} \int {\cal D} (A, H, c,
\bar{c}) \ e^{-S(A,H)-S_g + J \cdot A + K\cdot H + \bar{\chi}c + \chi
\bar{c}} \eeq 

\noi with

\beq
\label{2.6}
S(A,H) = \int d^4 x \left ( {1 \over 2} \left ( F_{\mu \nu}^a \right )^2
- {1 \over 2} F_{\mu \nu}^a \ H_{\mu \nu}^a + {1 \over 4} \left ( H_{\mu
\nu}^a \right )^2 \right ) \quad . \eeq

\noi Note that one could perform the Gaussian integral over $H$ in
eq. (\ref{2.5}) and obtain an equivalent formulation

\beq
\label{2.7}
e^{-G(J, K, \chi , \bar{\chi})} = {1 \over N} \int {\cal D} (A, c,
\bar{c}) \ e^{- S_{YM} - S_g + J \cdot A + K \cdot F + K\cdot K +
\bar{\chi}c + \chi \bar{c}} \eeq  

\noi The effective action including the auxiliary field $H$ is defined
through the Legendre transform 

\beq
\label{2.8}
\Gamma (A, H, c, \bar{c}) = G (J, K, \chi , \bar{\chi}) + J \cdot A + K
\cdot H + \bar{\chi} \cdot c + \chi \cdot \bar{c} \quad . \eeq

\noi A general effective action contains an infinite number of terms
with arbitrary powers in the fields and derivatives, which are just
restricted by the need to satisfy the Slavnov-Taylor identities. \par

Let us now consider an expansion of the $A_{\mu}^a$ and $H_{\mu \nu}^a$
dependent part of $\Gamma$ in powers of fields and derivatives: 

\bea
\label{2.9}
 \Gamma (A, H) 
&=& {Z \over 4} \left ( F_{\mu \nu} \right )^2 - {n \over
2} F_{\mu \nu} \ H_{\mu \nu} + {m^2 \over 4} \left ( H_{\mu \nu} \right
)^2 \nn \\ 
&+& {h \over 2} \left ( D_{\mu} \widetilde{H}_{\mu \nu} \right )^2 +
{\beta \over 2} \left ( D_{\mu} \ H_{\mu \nu} \right )^2 + {1 \over 2
\alpha} \left ( \partial_{\mu} \ A_{\mu} \right )^2 + \dots \eea

\noi Here $\widetilde{H}_{\mu \nu}^a$ is defined by $\widetilde{H}_{\mu
\nu}^a = {1 \over 2} \varepsilon_{\mu \nu \rho \sigma} \ H_{\rho
\sigma}^a$, and the covariant derivative $D_{\mu}$, acting on fields
$\varphi^a$ in the adjoint representation of the gauge group, by

\beq
\label{2.10}
D_{\mu} \varphi^a = \partial_{\mu} \varphi^a + \bar{g} \ f_{abc} \ 
A_{\mu} ^b \ \varphi^c \quad . \eeq

\noi In this approximation $\Gamma (A, H)$ thus depends on 7 parameters
$Z$, $n$, $m$, $h$, $\beta$, $\bar{g}$ and $\alpha$, which have to be
computed within some non-perturbative scheme. In the next section we
will discuss the Wilsonian exact renormalization group approach, but
here we proceed by discussing several important properties of $\Gamma
(A, H)$. \par 

First, the ``reduced'' gauge coupling $\bar{g}$ appearing in the
covariant derivative (\ref{2.10}) and in the non-abelian part of the
field strength $F_{\mu \nu}^a$ has no direct physical meaning. In order
to define a physical gauge coupling one first has to eliminate $H_{\mu
\nu}^a$ from eq. (\ref{2.9}) by its equations of motion. One obtains

\beq
\label{2.11}
\Gamma (A) = {Z_{eff} \over 4} \left ( F_{\mu \nu} \right )^2 + \dots     
\eeq

\noi where the dots denote terms of higher order in the covariant
derivatives which are induced by the terms $\sim h$, $\beta$. $Z_{eff}$
in (\ref{2.11}) is given by 

\beq
\label{2.12}
Z_{eff} = Z - {n^2 \over m^2} \quad .  
\eeq

\noi A physical gauge coupling (independent of field redefinitions of
$A_{\mu}^a$) is now given by

\beq
\label{2.13}
g_{phys} = \bar{g}/\sqrt{Z_{eff}} \quad .
\eeq 

Next we study the two point functions (in momentum space) of the fields
$A_{\mu}^a$ and $H_{\mu \nu}^a$, as obtained from the effective action
(\ref{2.9}) in the Landau gauge $\alpha \to 0$. One finds

\bminiG{}
\label{2.14a}
\left ( {\delta^2 \Gamma \over \delta \varphi_i (-p) \delta
\varphi_j(p)} \right )^{-1}_{A_{\mu}^a , A_{\nu}^b} 
&=& \delta_{ab} \left ( \delta_{\mu \nu} - {p_{\mu} p_{\nu} 
\over p^2} \right ) \cdot P_A(p^2) \quad , \nn \\
P_A(p^2) &=& {p^2 \beta + m^2 \over p^2 (Z m^2 - n^2) + Z \beta p^4}
\quad ,  \eeeq  
\beeq
\label{2.14b}
\left ( {\delta^2 \Gamma \over \delta \varphi_i (-p) \delta
\varphi_j(p)} \right )^{-1}_{A_{\mu}^a, H_{\rho \sigma}^b} 
&=& - i \delta_{ab} \left ( p_{\rho} \delta_{\mu \sigma} - p_{\sigma}
\delta_{\mu \rho} \right ) \cdot P_{AH}(p^2) \nn \\ 
P_{AH}(p^2) &=& {n \over p^2(Zm^2 - n^2) + Z \beta p^4} \quad ,  \eeeq 
\beeq
 \label{2.14c}
&&\left ( {\delta^2 \Gamma \over \delta \varphi_i (-p) \delta \varphi_j
(p)} \right )^{-1}_{H_{\rho \sigma}^a,H_{\kappa \lambda}^b} =
\delta_{ab} \left ( \delta_{\rho \kappa} \ \delta_{\sigma \lambda} -
\delta_{\rho\lambda} \ \delta_{\sigma \kappa} \right ) P_{HH,1}(p^2) \nn
\\ 
&&+ \delta_{ab} \left ( \delta_{\rho \kappa} \ p_{\sigma} \ p_{\lambda}
- \delta_{\rho \lambda} \ p_{\sigma} \ p_{\kappa} + \delta_{\sigma
\lambda} \ p_{\rho} \ p_{\kappa} - \delta_{\sigma \kappa} \ p_{\rho} \
p_{\lambda} \right ) P_{HH,2}(p^2) \quad , \nn \\ 
&&P_{HH,1}(p^2) = {1 \over hp^2 + m^2} \quad , \nn \\
&&P_{HH,2}(p^2) = {Z(h - \beta ) p^2 + n^2 \over \left ( hp^2 + m^2
\right ) \left ( p^2 \left ( Zm^2 - n^2 \right ) + Z\beta p^4 \right )}
\quad .  
\emini

\noi Let us first check the gluon propagator $P_A$ in the case where
$\Gamma (A, H)$ of eq. (\ref{2.9}) corresponds to the classical action
$S(A, H)$ of eq. (\ref{2.6}). After rescaling $H \to \Lambda H$ in
eq. (\ref{2.6}) in order to give it the appropriate dimension of a
bosonic field (with $\Lambda$ equal to, e.g., an UV cutoff) this choice
of $\Gamma (A, H)$ corresponds to the following choice of its parameters
$Z$, $n$, $m$, $h$, $\beta$ and $\bar{g}$:

\bea
\label{2.15}
Z &=& 2 \quad , \nn \\
n \ = \ m &=& \Lambda \quad , \nn \\
h \ = \ \beta &=& 0 \quad , \nn \\
\bar{g} &=& g_0 \quad .
\eea

\noi Then one finds that $P_A(p^2)$ is given by $1/p^2$ (and $Z_{eff}$
by 1) as it should be. \par

On the other hand, $P_A(p^2)$ enjoys a remarkable property if the three
parameters $Z$, $n$ and $m$ are related such that

\beq
\label{2.16}
Z_{eff} = Z - {n^2 \over m^2} \to 0 \quad .
\eeq

\noi Then the gluon propagator $P_A(p^2)$ becomes 

\beq
\label{2.17}
P_A(p^2) = {p^2 + \Lambda_c^2 \over Zp^4} \qquad , \qquad \Lambda_c^2 =
{m^2 \over \beta} \quad , \eeq

\noi i.e. for $p^2 << \Lambda_c^2$ the gluon propagator behaves like
$1/p^4$. In \cite{10r} we have emphasized that at the same time when
eq. (\ref{2.16}) is satisfied, the abelian projection of the action
(\ref{2.9}) allows for an explicit duality transformation, where the
dual action corresponds to the one of an abelian Higgs model in the
broken phase. The abelian projection corresponds simply to a vanishing
of the reduced gauge coupling $\bar{g}$. (In this case the effective
action would be invariant under a new gauge symmetry of the form $\delta
A_{\mu}^a = \Lambda_{\mu}^a$, $\delta H_{\mu \nu}^a =
\partial_{[\mu}A_{\nu ]}^a$, were it not for the ``gauge fixing term''
$\beta /2(\partial_{\mu} H_{\mu \nu})^2$ in eq. (\ref{2.9}).) \par

The fields appearing in the dual action are a dual abelian gauge field
$B_{\mu}$, a Goldstone boson $\varphi$ and a free massless scalar $\chi$
(whose origin can be traced back to the ``gauge fixing term'' $\sim
\beta$). The duality transformation is given by  

\bea
\label{2.18}
\sqrt{Z} F_{\mu \nu}^B &=& Z \ \widetilde{F}_{\mu \nu} - n
\widetilde{H}_{\mu \nu} \quad , \nn \\
\partial_{\nu} \varphi - n B_{\nu}/\sqrt{Zh} &=& \sqrt{h} \
\partial_{\mu} \ \widetilde{H}_{\mu \nu} \quad ,\nn \\
\partial_{\nu} \ \chi &=& \sqrt{\beta} \ \partial_{\mu} \ H_{\mu \nu}
\quad .   \eea 

\noi Assuming the relation (\ref{2.16}), the dual action
$\widetilde{\Gamma}$ becomes  

\beq
\label{2.19}
\widetilde{\Gamma}(B, \varphi ) = {1 \over 4} \left ( F_{\mu \nu}^B
\right )^2 + {1 \over 2} \left ( \partial_{\mu} \varphi - \widetilde{m}
B_{\mu} \right )^2 + {1 \over 2} \left ( \partial_{\mu} \chi \right )^2
\quad . \eeq 

\noi The mass $\widetilde{m}$ of the dual gauge field is given by

\beq
\label{2.20}
\widetilde{m}^2 = {n^2 \over Zh} = {m^2 \over h}
\eeq   

\noi where we assumed again (\ref{2.16}) to hold.  \par

Now we turn from previously published results to a new quantity of
interest, the correlator of two field strengths $F_{\mu \nu}^a$. Due to
the presence of a source $K_{\mu \nu}^a$ for the field strength in the
form (\ref{2.7}) for the partition function this correlator can simply
be expressed in terms of the functional $G$: 

\bea
\label{2.21}
\langle F_{\mu \nu}^a (x) \ F_{\rho \sigma}^b(0) \rangle &=& \left [ -
\delta_{ab} \left ( \delta_{\mu \rho} \ \delta_{\nu \sigma} -
\delta_{\mu \sigma} \ \delta_{\nu \rho} \right ) \delta^4(x) + {\delta G
\over \delta K_{\mu \nu}^a (x)} \ {\delta 
G \over \delta K_{\rho \sigma}^b(0)} \right. \nn \\
&-& \left . {\delta^2 G \over \delta K_{\mu \nu}^a (x) \ \delta K_{\rho
\sigma}^b (0)} \right ]_{K=J=0}  \quad . \eea

\noi (Here we have omitted the Schwinger strings, which we assume not to
significantly affect its $x$ dependence.) \par

After the Legendre transform (\ref{2.8}) the correlator becomes in terms
of the effective action $\Gamma$

\bea
\label{2.22}
\langle F_{\mu \nu}^a (x) \ F_{\rho \sigma}^b(0) \rangle &=& \left [ -
\delta_{ab} \left ( \delta_{\mu \rho} \ \delta_{\nu \sigma} -
\delta_{\mu \sigma} \ \delta_{\nu \rho} \right ) \delta^4(x) + H_{\mu
\nu}^a(x) \ H_{\rho \sigma}^b(0) \right. \nn \\ 
&+&\left .  \left ( {\delta^2 \Gamma \over \delta \varphi_i \ \delta
\varphi_j} \right )^{-1}_{H_{\mu \nu}^a(x) H_{\rho \sigma}^b(0)} \right
]_{{\delta \Gamma \over \delta H_{\mu \nu}^a} = {\delta \Gamma \over
\delta A_{\mu}^a} = 0} \quad . \eea 

\noi Assuming no vevs of the fields $H_{\mu \nu}^a$ and $A_{\mu}^a$,

\beq
\label{2.23}
H_{\mu \nu}^a = A_{\mu}^a = 0 \qquad \hbox{for} \qquad {\delta \Gamma
\over \delta H_{\mu \nu}^a} = {\delta \Gamma \over \delta A_{\mu}^a} = 0
\quad ,  \eeq 

\noi the correlator is thus given by the Fourier transforms of the
propagators given in eq. (\ref{2.14c}). Conventionally
\citt{6r}{8r}{11r} the correlator is decomposed into two Lorentz
invariant functions $D(x^2)$ and $D_1(x^2)$:  

\bea
\label{2.24}
&&\langle F_{\mu \nu}^a (x) \ F_{\rho \sigma}^b(0) \rangle = \Big [ -  
\delta_{ab} \left ( \delta_{\mu \rho} \ \delta_{\nu \sigma} -
\delta_{\mu \sigma} \ \delta_{\nu \rho} \right ) D(x^2) \nn \\ 
&&+ {1 \over 2} \left ( {\partial \over \partial x_{\mu}} \left (
x_{\rho} \ \delta_{\nu \sigma} - x_{\sigma} \ \delta_{\nu \rho} \right )
+ {\partial \over \partial x_{\nu}} \left ( x_{\sigma} \ \delta_{\mu
\rho} - x_{\rho} \ \delta_{\mu \sigma} \right ) \right ) D_1 (x^2) \Big ]
\eea 

\noi After taking the rescaling of $H$ by $\Lambda$ into account, our
expressions for the functions $D(x^2)$ and $D_1(x^2)$ become after
comparing (\ref{2.24}) with (\ref{2.14c}):

\bea
\label{2.25}
&&D(x^2) = \Lambda^2 \int {d^4p \over (2 \pi )^4} \ e^{ipx} \
P_{HH,1}(p^2) - \delta^4(x) \quad , \nn \\
&&D_1(x^2) = - 4 \Lambda^2 {d \over dx^2} \int {d^4p \over (2 \pi)^4} \
e^{ipx} \ P_{HH,2}(p^2)  \eea 

\noi with $P_{HH,1}$ and $P_{HH,2}$ as in eq. (\ref{2.14c}). Both
functions $D(x^2)$ and $D_1(x^2)$ have been measured on the lattice
\cite{11r}, and at least $D(x^2)$ is well fitted by a decaying 
exponential

\beq
\label{2.26}
D(x^2) \sim e^{-M |x|} \quad , \quad M \sim 1 \ \hbox{GeV} \quad ,
\eeq

\noi for $|x| \ \lsim \ 1$ fm. \par

Subsequently we concentrate on the function $D(x^2)$, whose
non-vanishing is known to be responsible for the area law of the Wilson
loop, if the cluster expansion converges \cite{6r}. Note that in our
case, if $\Gamma$ would correspond to the classical action $S$ and hence
$h = 0$, $m^2 = \Lambda^2$, the propagator $P_{HH,1}$ is simply
$1/\Lambda^2$ and thus $D(x^2)$ vanishes, since two $\delta$-functions
cancel each other in eq. (\ref{2.25}). For a general action (\ref{2.9})
$D(x^2)$ is easily evaluated and one finds, for $|x| \not= 0$,

\beq
\label{2.27}
D(x^2) = {m \Lambda^2 \over 4 \pi^2 h^{3/2} |x|} K_1 \left ( {xm \over 
\sqrt{h}} \right ) \eeq

\noi where $K_1$ is a Bessel function. Thus one obtains indeed an
exponential decay (modulo powers) as in (\ref{2.26}), with

\beq
\label{2.28}
M = {m \over \sqrt{h}} \quad .
\eeq 

\noi The result (\ref{2.27}) agrees with the one obtained in \cite{8r}
on the basis of a dual Abelian Higgs model, and in \cite{9r} from a
confining string theory. In all cases the inverse exponential decay
length $M$, eq. (\ref{2.28}), coincides with the mass $\widetilde{m}$
of the dual gauge field, cf. the second equality in eq. (\ref{2.20}). 
\par

Since our effective action (\ref{2.9}) contains also the gluon field we
can investigate, in how far the behaviour of $D(x^2)$ is related to the
gluon propagator. Surprisingly there is no direct relation: a $1/p^4$
behaviour of the gluon propagator as in eq. (\ref{2.17}) depends
crucially on the relation (\ref{2.16}) to hold, i.e. on a relation
between the parameters $Z$, $n$, and $m$. On the other hand, the result
(\ref{2.27}) does not at all depend on a relation like
eq. (\ref{2.16}). Even if this relation holds, the parameter $M$ in
(\ref{2.28}) is not directly related to the parameter $\Lambda_c$
characterizing the $1/p^4$ behaviour of the gluon propagator, unless the
parameters $h$ and $\beta$ in the effective action (\ref{2.9}) happen
to be close to each other. (Actually, the function $D_1(x^2)$ does
depend on the parameters $Z$, $n$ and $m$. However, present lattice data
\cite{11r} do not yet allow to study this function in detail.) Thus we
see  that the different pictures of confinement -- based on the gluon
propagator or on the field strength correlator -- are not directly
related. \par 

Of course it can be argued that the result (\ref{2.27}) is only a
trivial consequence of our ansatz (\ref{2.9}) for the effective action,
whose parameters should eventually be computed from the Yang Mills
Lagrangian. Note, however, that the action (\ref{2.9}) implies the
result (\ref{2.27}) only if the parameters $h$ and $m^2$ turn out to be
non-zero, finite and positive. The result of computation of the
parameters of the action within the context of the Wilsonian exact
renormalization group approach will be presented in the next section. 

\mysection{An Infrared Fixed Point of Exact Renormalization Group
Equations for Yang Mills Theories} 
\hspace*{\parindent}

In the recent years much progress has been made in applying exact
renormalization group 
equations \cite{14r} to gauge theories \citr{12r}{13r}{15r}{17r}. Also
sources coupled to composite fields can be introduced in this formalism
\cite{18r} which allows to apply it to the present case. \par

The exact renormalization group approach requires the introduction of an
``artificial'' infrared cutoff $k$ into the partition function
(\ref{2.1}). Then one exploits the facts that the corresponding $k$
dependent effective action $\Gamma_k$ becomes equal to the classical
action $S$ in the limit $k \to \infty$ (up to additional terms
determined by the modified Slavnov-Taylor identities \citd{15r}{16r}),
and that an exact functional differential equation fixing the $k$
dependence of $\Gamma_k$ can be derived. Integrating this Wilsonian
exact renormalization group equation from some large value $k = \Lambda$
down to $k = 0$ provides us with the physical effective action
$\Gamma_{k=0}$ in terms of the parameters of some ``high energy''
effective action $\Gamma_{\Lambda}$. By construction no ultraviolet
invergences appear in this approach, if both $\Lambda$ and
$\Gamma_{\Lambda}$ are assumed to be finite. \par 

To be concrete, in the present case with a source $K$ coupled to the
field strength $F$, one defines the functional $G_k(J,K, \chi ,
\bar{\chi})$ including the infrared cutoff $k$ by a corresponding
modification of eq. (\ref{2.7}): 

\beq
\label{3.1}
e^{-G_k(J,K, \chi , \bar{\chi})} = {1 \over N} \int {\cal D}(A, c ,
\bar{c}) \ e^{-S_{YM} - S_g - \Delta S_k + J \cdot A + K \cdot F + K
\cdot K + \bar{\chi} \cdot c + \chi \cdot \bar{c}} \eeq

\noi where $\Delta S_k$ implements the infrared cutoff for the gauge and
ghost fields: 

\beq
\label{3.2}
\Delta S_k = \int {d^4p \over (2 \pi )^4} \left [ {1 \over 2}
A_{\mu}^a(-p) \ R_{\mu \nu}^k(p^2) \ A_{\nu}^a (p) + \bar{c}^a(-p) \
R_g^k(p^2) c^a(p) \right ] \quad . \eeq 

\noi The functions $R_{\mu \nu}^k$ and $R_g^k$ modify the gauge and
ghost propagators such that modes with $p^2 << k^2$ are
suppressed. Convenient choices are 

\bea
\label{3.3}
&&R_{\mu \nu}^k(p^2) = \left ( p^2 \ \delta_{\mu \nu} + \left ( {1 \over
\alpha} - 1 \right ) p_{\mu} \ p_{\nu} \right ) {e^{-p^2/k^2} \over 1 -
e^{-p^2/k^2}} \quad , \nn \\ 
&&R_g^k(p^2) = p^2 {e^{-p^2/k^2} \over 1 - e^{-p^2/k^2}} \quad . \eea

\noi The effective action in the presence of the infrared cutoff $k$ is
again defined through the Legendre transform

\beq
\label{3.4}
\widetilde{\Gamma}_k(A, H, c, \bar{c}) = G_k (J, K, \chi , \bar{\chi}) + 
J \cdot A + K \cdot H + \bar{\chi} \cdot c + \chi \cdot \bar{c} \quad
. \eeq 

\noi Some expressions become more handsome when written in terms of
$\Gamma_k$ which is given by $\widetilde{\Gamma}_k$ with the infrared
cutoff term $\Delta S_k$ subtracted:

\beq
\label{3.5}
\Gamma_k (A, H, c, \bar{c}) = \widetilde{\Gamma}_k (A, H, c, \bar{c}) -
\Delta S_k \quad . \eeq

\noi From the path integral (\ref{3.1}) and the Legendre transform
(\ref{3.2}) it is straightforward to derive the exact renormalization
group equations \citm{12r}{18r} 

\beq
\label{3.6}
\partial_k \Gamma_k = {1 \over 2} \int {d^4p \over (2 \pi )^4} \
\partial_k \ R^k(p^2)_{ij} \left ( {\delta^2 \widetilde{\Gamma}_k \over
\delta \bar{\varphi}_{\ell} (-p) \ \delta \varphi_m (p)} \right
)^{-1}_{ji} \quad . \eeq 

\noi Here the fields $\varphi_i \equiv ( A_{\mu}^a, H_{\mu \nu}^a, c^a,
\bar{c}^a)$ denote all possible fields appearing as arguments of
$\Gamma_k$ or $\widetilde{\Gamma}_k$, and the index $i$ corresponds to
the field type and Lorentz and gauge group indices. The matrix
$R_{ij}^k$ has non-vanishing matrix elements only in the subsectors
$(A_{\mu}^a, A_{\nu}^a)$ and $(\bar{c}^a, c^a)$. The inverse functional 
$(\delta^2\widetilde{\Gamma}_k/\delta \bar{\varphi}_{\ell} \delta
\bar{\varphi}_m)_{ji}^{-1}$, however, has to be constructed on the
complete space spanned by $(A_{\mu}, H_{\mu \nu}, c, \bar{c})$ including
the auxiliary field $H_{\mu \nu}$. \par

The right-hand side of eq. (\ref{3.6}) corresponds to a one loop diagram
with an arbitrary number of vertices or external lines, and an insertion
of $\partial_k R_{\mu \nu}^k$ or $\partial_k R_g^k$ into a gauge field
or ghost propagator. The vertices and propagators have to be derived
from the $k$ dependent effective action $\widetilde{\Gamma}_k$. For a
given parametrization of $\Gamma_k$ (or $\widetilde{\Gamma}_k$),
non-linear differential equations for the $k$ dependence of the
corresponding parameters are obtained from eq. (\ref{3.6}) by comparing
equal powers of fields and/or derivatives acting on the fields on both
sides.  \par 

Now we turn to a parametrization of the $A_{\mu}$ and $H_{\mu \nu}$
dependent part of $\Gamma_k$ in the form of eq. (\ref{2.9}), where from
now on the parameters $Z$, $n$, $m$, $h$ and $\beta$ depend on the
infrared cutoff $k$. Differential equations describing the $k$
dependence of these parameters are obtained from eq. (\ref{3.6}) by
considering terms quadratic in $A_{\mu}$ or $H_{\mu \nu}$, and they 
have been derived in the Landau gauge in ref. \cite{10r}. \par

On the right-hand side of these differential equations appears the
``reduced'' gauge coupling $\bar{g}$, since all vertices (either from
the non-abelian part of $F_{\mu \nu}^a$ or from the covariant
derivatives $D_{\mu}$) are proportional to $\bar{g}$. Thus the
renormalization group equation describing the $k$ dependence of
$\bar{g}$ is also needed. \par

In principle this renormalization group equation could be obtained from
the three gluon vertex or the terms trilinear in $A_{\mu}$ in
eq. (\ref{3.6}), but due to the large number of contributing diagrams it
is much more convenient to introduce the ghost sector and to consider
the relations implied by the Slavnov-Taylor identities. \par

The simplest non-trivial parametrization of the ghost sector of
$\Gamma_k$ is given by

\bea
\label{3.7}
\Gamma_k^{ghost} &=& Z_g \ \partial_{\mu} \ \bar{c} \ D_{\mu} \ c \nn \\
&\equiv & Z_g \ \partial_{\mu} \ \bar{c}^a \ \partial_{\mu} \ c^a + Z_g \ 
\bar{g} \ f_{abc} \ \partial_{\mu} \ \bar{c}^a \ A_{\mu}^b \ c^c \quad
. \eea 

Note that the Slavnov-Taylor identities imply that the reduced coupling
$\bar{g}$ in eq. (\ref{3.7}) equals the coupling $\bar{g}$ implicit in
eq. (\ref{2.9}). (Here we neglect modifications of the Slavnov-Taylor
identities for $k \not= 0$ \citd{15r}{16r} which vanish for $k \to
0$). Furthermore the ghost-gluon coupling receives no quantum
corrections in the Landau gauge. In terms of the exact renormalization
group equations this implies $\partial_k(Z_g \bar{g}) = 0$ or 

\beq
\label{3.8}
\bar{g}^{-1} \ \partial_k \ \bar{g} = - Z_g^{-1} \ \partial_k \ Z_g
\quad . \eeq

Thus the $k$ dependence of $\bar{g}$ can be obtained from the $k$
dependence of the ghost wave function normalization $Z_g$, which is much
easier to compute. Within the present parametrization of $\Gamma_k$ the
corresponding exact renormalization group equation has also been given
in \cite{10r}, hence a closed set of differential equations describing
the $k$ dependence of the 6 parameters $Z$, $n$, $m$, $h$, $\beta$ and
$\bar{g}$ has been obtained. They have been integrated numerically, with
boundary conditions such that at $k = \Lambda$ $\Gamma_k (A, H, c,
\bar{c})$ corresponds to the classical action (cf. eq. (\ref{2.15})): 

\bea
\label{3.9}
Z(\Lambda ) &=& 2 \quad , \qquad n(\Lambda ) = m (\Lambda ) = \Lambda
\quad , \nn \\ 
h(\Lambda ) = \beta (\Lambda ) &=& 0 \quad , \qquad \bar{g} (\Lambda ) =
g_0 \quad , \nn \\
Z_g(\Lambda ) &=& 1 \quad . \eea 
As a result of the renormalization group flow we found indeed that for
small $k$ $Z_{eff}(k)$ (defined in terms of $Z$, $n$ and $m$ in
eqs. (\ref{2.12}) and (\ref{2.16})) vanishes. However, it was not
possible to reach $k = 0$, since a Landau singularity appeared in
$\bar{g}(k)$ at some small, but finite value of $k = k_c$: within the
present approximation of $\Gamma_k$ the right-hand side of the equation
for $\partial_k \bar{g}(k)$ is negative definite and proportional to
$\bar{g}^2$ (as in the case of one loop $\beta$ function for Yang Mills
theories), thus a Landau singularity cannot be avoided. Clearly, this is
an artifact of the neglect of the contributions of higher dimensional
operators to the right-hand side of the exact renormalization group
equations. \par

Let us now go beyond the previous simple truncation of $\Gamma_k$ and
consider the effects of such higher dimensional operators. Since the
running of $\bar{g}(k)$ is governed by the running of $Z_g(k)$,
cf. eq. (\ref{3.8}), we have to consider such contributions to the
right-hand side of the equation for $\partial_k Z_g(k)$. First,
operators with the same powers of fields as in eq. (\ref{3.7}), but
involving higher derivatives, will not solve the problem: these have
already been considered in the second of refs. \cite{13r}, and they 
do not modify the negative definiteness of $\partial_k \bar{g}(k)$ (or
the positive definiteness of $\partial_k Z_g (k)$, implying $Z_g (k_c) =
0$ with $k_c$ finite). \par

Thus we proceed by adding an operator of higher powers in the fields to the
ghost sector of $\Gamma_k$, which will contribute to the running of $Z_g(k)$.
We replace $\Gamma_k^{ghost}$ of eq. (\ref{3.7}) by

\bea
\label{3.10}
\Gamma_k^{ghost} &=& Z_g \ \partial_{\mu} \ \bar{c} \ D_{\mu} \ c +
\lambda \ \partial_{\mu} \ \bar{c} \ D_{\mu} \left ( c \ F_{\nu \rho} \
F_{\nu \rho} \right ) \nn \\
&\equiv & Z_g \ \partial_{\mu} \ \bar{c}^a \ \partial_{\mu} \ c^a + Z_g
\ \bar{g} \ f_{abc} \ \partial_{\mu} \bar{c}^a \ A_{\mu}^b \ c^c \nn \\
&+& \lambda \ \partial_{\mu} \ \bar{c}^a \ \partial_{\mu} \left ( c^a \
F_{\nu \rho}^d \ F_{\nu \rho}^d \right ) + \lambda \ \bar{g} \ f_{abc} \
\partial_{\mu} \bar{c}^a \ A_{\mu}^b \ c^c \ F_{\nu \rho}^d \ F_{\nu
\rho} ^d \quad . \eea 

\noi The action of the covariant derivative $D_{\mu}$ in the term $\sim 
\lambda$ in eq. (\ref{3.10}) has been chosen such that the
Slavnov-Taylor identities are still satisfied in a simple way:
technically speaking, since the variation of $\Gamma_k^{ghost}$ with
respect to $\partial_{\mu} \bar{c}^a$ is still proportional to a total
covariant derivative, the effective BRST variation of the gluon field
$A_{\mu}^a$ is still a total covariant derivative as before. \par

In principle, also a term quartic in the ghost fields could have been
added to $\Gamma_k^{ghost}$. However, the corresponding contribution to
$\partial_k Z_g$ involves only a diagram with an internal ghost
propagator. This contribution is relatively small compared to the
contribution obtained below in the limit where the gluon propagator
becomes infrared singular, cf. our results below. \par

After inserting $\Gamma_k^{ghost}$ of eq. (\ref{3.10}) into the exact
renormalization group equation (\ref{3.6}) and expanding the right-hand
side to the order $\partial_{\mu} \bar{c} \ \partial_{\mu} c$ one
obtains the contributions to $\partial_k Z_g$ which are shown
diagrammatically in fig. 1. The ``tadpole'' diagram involving a
ghost-ghost-gluon-gluon vertex is proportional to the coupling $\lambda$
in eq. (\ref{3.10}). \par 

A priori a huge number of diagrams contribute to the running of
$\partial_k \lambda$, alone 15 to the order $\lambda^0$. Since we are
not (yet) interested in precise quantitative results, but rather in the
essential features of the system of differential equations, we will only
take the leading contributions into account: to the orders $\lambda^1$
and $\lambda^2$ (no higher orders in $\lambda$ exist) we consider only
those diagrams, which are leading in the case of an infrared singular
gluon propagator, i.e. for $Z_{eff} \to 0$. \par

Instead of including all the contributions to $\partial_k \lambda$ to
the order $\lambda^0$, we will imitate these contributions by a small,
but non-vanishing value of $\lambda$ at the starting point $k =
\Lambda$. This procedure is justified, since one finds that for $k \to
0$ $\lambda$ becomes extremely large, and all contributions $\sim
\lambda^0$ to $\partial_k\lambda$ become relatively negligeable. Also we
have verified that the numerical results for $k \to 0$ are practically
independent of the starting point value of $\lambda$ provided it is
small enough. This is just a manifestation of universality of the
Wilsonian exact renormalization group flow, i.e. generally the results
for $k \to 0$ depend only very weakly on the irrelevant couplings in
$\Gamma_k$ at the starting point $k = \Lambda$. Hence the diagrams which
contribute to $\partial_k \lambda$ are finally just those shown in
fig. 2, which are of the orders $\lambda^1$ and $\lambda^2$. \par

In order to write down the resulting exact renormalization group
equations for $\partial_k Z_g$ and $\partial_k \lambda$ we use the
following notations: we need the gluon propagator function $P_A(p^2)$ of
eq. (\ref{2.14a}), and in addition the ghost propagator $P_G(p^2)$ given
by 

\beq
\label{3.11}
P_G(p^2) = {1 \over Z_g p^2} \quad .
\eeq               

\noi The presence of the infrared cutoff requires the replacement of $Z$
by $Z + \widetilde{R}$ in eq. (\ref{2.14a}), and of $Z_g$ by $Z_g +
\widetilde{R}$ in eq. (\ref{3.11}) , with

\beq
\label{3.12}
\widetilde{R} = {e^{-p^2/k^2} \over 1 - e^{-p^2/k^2}}
\eeq  

\noi for the choices in eq. (\ref{3.3}). $g_0$ denotes the bare gauge
coupling (appearing at the ghost gluon vertex), and, following the
discussions below eq. (\ref{3.7}) the coupling $\bar{g}$ equals
$g_0/Z_g$. For a SU(N) gauge group we then obtain from figs. 1 and 2:

\beq
\label{3.13}
\partial_k Z_g = \int {q^2 dq^2 \over 16 \pi^2} \partial_k \
\widetilde{R}(q^2) \left [ {3Ng_0^2 \over 4} \ q^2 \left ( P_A^2 \ P_G +
P_A \ P_G^2 \right ) - 12 N \ \lambda \ q^4 \ P_A^2 \right ] \quad ,
\eeq

\beq
\label{3.14}
\partial_k \lambda = \int {q^2 dq^2 \over 16 \pi^2} \ \partial_k \
\widetilde{R}(q^2) \left [ - {9N {\bar{g}}^2 \over 2} \ \lambda \ Z^2 \
q^4 \ P_A^4 \ + 4 \lambda^2 \ q^6 \ P_A^2 \ P_G \right ] \quad . \eeq

The presence of the term $\sim \lambda$ in $\Gamma_k^{ghost}$ in
eq. (\ref{3.10}) affects also the exact renormalization group equation
for $\partial_k Z$, but not the equations of the remaining parameters
$n$, $m$, $h$ and $\beta$. For completeness we present these five
equations below. Their right-hand sides receive contributions from the
vertices from the non-abelian parts of $F_{\mu \nu}^a$, i.e. from the
first two terms in $\Gamma (A,H)$ of eq. (\ref{2.9}), and from the
vertices due to the non-abelian parts of the covariant derivatives
acting on $\widetilde{H}_{\mu \nu}^a$ or $H_{\mu \nu}^a$, which are thus
proportional to $h$ or $\beta$. (These latter contributions are not very
important numerically.) \par

Below $P_A$, $P_{AH}$ etc. denote the propagator functions of
eqs. (2.14), with the replacements of $Z$ by $Z + \widetilde{R}$ as
below eq. (\ref{3.11}). Because of the required development in powers of
derivatives (or external momenta) the quantities $p^2dP_A/dp^2$
etc. appear frequently, and we define for convenience 

\beq
\label{3.15}
P' \equiv p^2 {dP \over dp^2} \quad , \qquad P'' = p^4 {d^2P \over
(dp^2)^2} \eeq

\noi for all propagator functions $P_A$, $P_{AH}$ etc. With $\partial_k
\equiv k^2 d/dk^2$ and for a SU(N) gauge group the five exact
renormalization group equations are  

\bminiG{}
\label{3.16a}
\partial_k \ Z &=& {N\bar{g}^2 \over 16 \pi^2} \int_0^{\infty} dp^2 
\ p^4 \ \partial_k \ R^k \Big
[ Z^2 \left ( {31 \over 6} P_A^3 + 3P_A^2 P'_A + P_A^2 P''_A \right ) 
\nn \\
&&- n Z P_A \Big ( {16 \over 3} P_A P_{AH} + {8 \over 3} P_{AH}
 P'_A + P_{AH} P''_A 
 + {10 \over 3} P_A P'_{AH} + P_A P''_{AH} \Big ) \nn \\
&&+ n^2 \Big ( P_A \ P_{AH} \left ( {3 \over 2} P'_{AH} + {1 \over 2} 
P''_{AH} \right ) 
+ P^2_{AH} \left ( {5 \over 12} P_A + {7 \over 12} P'_A + {1 \over 4} 
P''_A \right ) \nn \\
&&+ P_A^2 \left ( p^{-2} \left ({9 \over 4} P'_{HH,1} + 
{13 \over 12} P''_{HH,1} \right ) + 
{3 \over 4} P_{HH,2} + {11
\over 12} P'_{HH,2} + {1 \over 4} P''_{HH,2} \right ) \Big ) \nn \\
&& + Z \beta P_A P_{AH} p^2 \left ( 8 P_{AH} + 8 P'_{AH} + 2 P''_{AH} 
\right ) \nn \\
&& + n \beta \ \big ( P^2_{AH} p^2 \left ( 2 P_{AH} - {11 \over 3} P'_{AH}
- P''_{AH} \right ) - P_A P_{AH} \left ( 6 P'_{HH,1} + {8 \over 3} P''_{HH,1}
\right ) \nn \\
&& - P_A P_{AH} p^2 \left ( 3 P_{HH,2} + {13 \over 3} P'_{HH,2} +
P''_{HH,2} \right ) \big ) \nn \\
&& + \beta^2 \big ( P^2_{AH} p^2 \left ( P_{HH,1} + {29 \over 6} P'_{HH,1} 
+ {11 \over 6} P''_{HH,1} \right ) \nn \\
&& + P^2_{AH} p^4 \left ( {13 \over 3} P_{HH,2} + 5 P'_{HH,2} + P''_{HH,2}
\right ) \big ) \nn \\
&& + n h P_A P_{AH} \left ( 5 P'_{HH,1} + {5 \over 3} P''_{HH,1} \right ) 
+ h^2 P^2_{AH} p^2 \left ( P_{HH,1} + {19 \over 6} P'_{HH,1} + 
{5 \over 6} P''_{HH,1} \right )\nn \\
&& + \beta h P^2_{AH} p^2 \left ( 5 P'_{HH,1} - {5 \over 3} P''_{HH,1}
\right ) 
+ {1 \over 6} Z_g^2 \ P_g^3 \Big ] \quad , \\
\label{3.16b} 
\partial_k \ n &=&
{N \bar{g}^2 \over 16 \pi^2} \int_0^{\infty} dp^2 \ p^4 \ \partial_k 
\ R^k \Big [ \ n \ P_A^2 \left ( {5 \over 2} Z \ P_A - {3 \over 2} n\ 
P_{AH} \right ) \nn \\
&& + n \beta P_A \left ( 6 p^2 P^2_{AH} + 3 P_A P_{HH,1} + 
{3 \over 2} p^2 P_A P_{HH,2} \right ) \nn \\
&& - n h P^2_A \left ( {3 \over 2} P_{HH,1} + P'_{HH,1} \right ) 
- 8 Z \beta p^2 P^2_A P_{AH} \nn \\
&& - \beta^2 P_{AH} p^2 \left ( {9 \over 2} P_A P_{HH,1} + 
3 p^2 P^2_{AH} + 3 p^2 P_A P_{HH,2} \right ) 
\nn \\ 
&& - h^2 p^2 P_A P_{AH} \left ({3 \over 2} P_{HH,1} + P'_{HH,1} \right ) 
\nn \\
&& + \beta h p^2 P_A P_{AH} \left ( 3 P_{HH,1} + P'_{HH,1} \right ) 
\Big ],  \\
\label{3.16c}
\partial_k \ m^2 &=& {N \bar{g}^2 \over 16 \pi^2} \int_0^{\infty} dp^2 \ 
p^4 \ \partial_k \ R^k \Big[ n^2 P_A^3 
- 4 \beta n p^2 P^2_A P_{AH} - 3 (\beta + h) P_A^2 \nn \\
&& + \beta^2 P_A \left ( 2 p^4 P_A P^2_{AH} + p^2 P_A P_{HH,1} + p^4
P_A P_{HH,2} \right ) \nn \\
&& + h^2 p^2 P^2_A P_{HH,1} \Big ]
\quad , \\
\label{3.16d}
\partial_k \ Z_g &=& {N\bar{g}^2 \over 16 \pi^2} \int_0^{\infty} dp^2 \ p^4 
\ \partial_k \ R^k \
Z_g^2 \ P_A \ P_g {3 \over 4} \left ( P_A + P_g \right ) \quad , \\
\label{3.16e}
\partial_k \ h &=& {N \bar{g}^2 \over 16 \pi^2} \int_0^{\infty} dp^2 \ p^4 
\ \partial_k \ R^k \Big [ \ n^2 \
P_A^2 \ p^{-2} \left ( {1 \over 6} P_A + {4 \over 3} P'_A + {2 \over 3} 
P''_A \right ) \nn \\
&& + \beta n \Big ( -P_A^2 \left ( {10 \over 3} P_{AH} + {16 \over 3}
P'_{AH} + {4 \over 3} P''_{AH} \right ) 
- P_A P_{AH} \left ( {8 \over 3} P'_A + {4 \over 3} P''_A \right )
\Big ) \nn \\ 
&& + h n P_A \left ( 2 P_A P_{AH} + 2 P_A P'{AH} - 2 P'_A
P_{AH} \right ) \nn \\
&& + \beta^2 \Big ( P^2_A \left ( {3 \over 2} P_{HH,1} + {8 \over 3}
P'_{HH,1} + {2 \over 3} P''_{HH,1} \right ) 
+ P^2_{AH} p^2 \left ( {5 \over 3} P_A + {4 \over 3} P'_A + {2 \over 3}
P''_A \right ) \nn \\
&& + P_A P_{AH} p^2 \left ( {8 \over 3} P'_{AH} + {2 \over 3} P''_{AH} 
\right ) 
+ P_A^2 p^2 \left ( {25 \over 6} P_{HH,2} + 4 P'_{HH,2} + {2 \over 3}
P''_{HH,2} \right ) \Big ) \nn \\
&& + h^2 P^2_A \left ({9 \over 2} P_{HH,1} + {4 \over 3} P'_{HH,1} +
{1 \over 3} P''_{HH,1} 
 + {3 \over 2} p^2 P_{HH,2} \right ) \nn \\
&& + \beta h \Big ( - P^2_A \Big ( 3 P_{HH,1} + 2 P'_{HH,1} \Big )
+ P^2_{AH} p^2 \Big ( P_A 
+ 2 P'_A \Big ) \nn \\ 
&& - P^2_A p^2 \Big ( 5 P_{HH,2} + 2 P'_{HH,2} \Big )
\Big ) \Big ]
\quad , \\ 
\label{3.16f}
\partial_k \ \beta &=& {N \bar{g}^2 \over 16 \pi^2} \int_0^{\infty} dp^2
\ p^4  \ \partial_k \ R^k \Big [ \ n^2 \
P_A^2 p^{-2} \left ( {1 \over 12} P_A + {2 \over 3} P'_A + {1 \over 3}
P''_A  \right ) \nn \\
&& + \beta n \Big ( -P_A^2 \left ( {14 \over 3} P_{AH} + {8 \over 3}
P'_{AH} + {2 \over 3} P''_{AH} \right ) 
- P_A P_{AH} \left ( {4 \over 3} P'_A + {2 \over 3} P''_A \right )
\Big ) \nn \\ 
&& + \beta^2 \Big ( P^2_A \left ( {9 \over 2} P_{HH,1} + {4 \over 3}
P'_{HH,1} + {1 \over 3} P''_{HH,1} \right ) 
 + P^2_{AH} p^2 \left ( {16 \over 3} P_A + {2 \over 3} P'_A + {1 \over 3}
P''_A \right ) \nn \\
&& + P_A^2 p^2 \left ( {13 \over 3} P_{HH,2} + 2 P'_{HH,2} + {1 \over 3}
P''_{HH,2} \right ) 
+ P_A P_{AH} p^2 \left ( {4 \over 3} P'_{AH} + {1 \over 3} P''_{AH} 
\right ) \Big ) \nn \\
&& + h^2 P^2_A \left ({3 \over 2} P_{HH,1} + {8 \over 3} P'_{HH,1} +
{2 \over 3} P''_{HH,1} \right ) 
 \nn \\ && 
- \beta h P^2_A \left ( 3 P_{HH,1} + 2 P'_{HH,1} \right )
 \Big ]. \emini

\noi Note that all integrals are trivially ultraviolet finite since,
with the present choice of $\widetilde{R}$, $\partial_k\widetilde{R}$
decreases exponentially for large $p^2$, and infrared finiteness is
ensured by the presence of the infrared cutoff terms in the
propagators. The integrals in eqs. (3.16), and hence the integration of
the exact renormalization group equations for all parameters $Z$, $n$,
$m$, $h$, $\beta$, $Z_g$  and $\lambda$ have to be performed
numerically. \par 

It has already been verified in \cite{10r} that near the starting point
$k = \Lambda$, with the parameters as in eq. (\ref{3.9}), the correct
one loop $\beta$ functions are obtained. Thus both $Z_{eff}$ as defined
in eq. (\ref{2.12}) and $Z_g$ decrease for decreasing $k$, i.e. the
reduced coupling $\bar{g} = g_0/Z_g$ increases. Also $\lambda$ increases
for decreasing $k$, because the first term on the right-hand side of
eq. (\ref{3.13}) dominates. \par

This trend continues for some orders of magnitude of decreasing $k$,
during which $Z_{eff}$ becomes extremely small. Then, at some critical
scale $k_c$, a violent transition occurs, which is entirely governed by
the two renormalization group equations (\ref{3.13}) and
(\ref{3.14}). (It is well known that coupled non-linear differential
equations can lead to quasi-singular solutions.) First, the right-hand 
side of eq. (\ref{3.13}) turns negative, since $\lambda$ continued to
increase. As a consequence $Z_g$ increases for decreasing $k$, i.e. the
reduced coupling $\bar{g}$ decreases. This decrease is very rapid;
nearly instantly $\bar{g}$ drops to a value very close to 0. At some
stage during this decrease, however, the second term on the right-hand
side of eq. (\ref{3.14}) starts to dominate. From this point onwards 
also $\lambda$ decreases for decreasing $k$. \par

Soon thereafter all evolution comes practically to an end, since all
right-hand sides of the renormalization group equations become
numerically tiny: the first terms are proportional to $1/Z_g$ with $Z_g$
extremely large (remember that both $\bar{g}$ and $P_G$ behave like
$1/Z_g$), and the second terms are suppressed by a power of $\lambda
k^4$ for $k \to 0$. \par 

In figs. 3, 4 and 5 we show our results, obtained numerically with 
$\bar{g}(\Lambda ) = g_0 = 1.2$ and $\lambda (\Lambda ) = 0.01$, for
$Z_{eff}(k)$, $\bar{g}(k) = g_0/Z_g(k)$ and $\lambda (k)$
respectively. First, in fig.~3, we plot $log (Z_{eff})$ versus $t = -
log (k^2/\Lambda ^2)$. The evolution is 
from left to right: for $k^2 = \Lambda^2$ $t$ is zero, and $t$ increases
for decreasing $k$. One recognizes the decrease of $Z_{eff}$ up to $k =
k_c$~; for $k$ below $k_c$ (or $t$ above $t_c$) the evolution of
$Z_{eff}$ stops: the right-hand side of all renormalization group
equations of $Z$, $n$, $m$, $h$ and $\beta$ are proportional to
$\bar{g}^2$, and for $t > t_c$ $\bar{g}^2$ is vanishingly small. \par 

This behaviour of $\bar{g}$ is shown in fig. 4, where we plot $log
(\bar{g})$ versus $t$. First, as described above, $\bar{g}$ increases
until it becomes nearly singular. Then, at $k = k_c$, it drops instantly
to a tiny value and remains there subsequently. \par

In fig. 5 we plot $log (\lambda )$ versus $t$. One observes its steady
increase, its maximum at $k = k_c$ and its subsequent drop to a somewhat
smaller value. \par

We have checked the independence of our result on the initial values of
$\bar{g}$ and $\lambda$ at the starting point $k = \Lambda$: we have
varied $\bar{g}(\Lambda ) = g_0$ between 1.0 and 1.5, and $\lambda
(\Lambda )$ between 0.1 and 0.001 and obtained always qualitatively the
same behaviour of $\bar{g}$, $\lambda$ and the other parameters, just
the scale $k_c$ varies correspondingly. \par

We have not been able to describe the quasi-singular behaviour of
$\bar{g}$ and $\lambda$ around the critical scale analytically. However,
this behaviour becomes already apparent within a simplified toy model:
it is possible to estimate the momentum integrals on the right-hand
sides of eqs. (\ref{3.13}) and (\ref{3.14}), where the main
contributions come from $q^2 \sim k^2 \cdot ( - log (Z_{eff}))$. Then
these equations can approximately be written in terms of $Z_{eff}$,
$\lambda$ and $Z_g$ (with $\bar{g} \sim 1/Z_g$, $P_G \sim 1/q^2 Z_g)$ as
follows: 

\bea
\label{3.17}
&&\partial_k \ Z_g = {c_1 \over Z_{eff} \ Z_g} - {c_2 \ \lambda \ k^4
\over Z_{eff}} \quad , \nn \\ 
&&\partial_k \ \lambda = {c_3 \ \lambda^2 \ k^4 \over Z_{eff} \ Z_g} -
{c_4 \ \lambda \over Z_{eff} \ Z_g^2} \eea

\noi with $c_1 \dots c_4$ of ${\cal O}(1)$. Already for constant
$Z_{eff}$ (note that $Z_{eff}$ varies only weakly around $k_c$) the
solutions for $Z_g$ and $\lambda$ of the simplified system (\ref{3.17})
of coupled non-linear differential equations have the same properties as
the ones shown in figs. 4 and 5. This constitutes an independent check
of our previous numerical results, and shows also that the qualitative
features are independent of additional small contributions from
neglected diagrams, i.e. the precise values of the constants 
$c_1 \dots c_4$. \par 

We can also insert the numerical values for $Z_{eff}$, $Z_g$, $\lambda$
and $k$ for $k < k_c$ into the right-hand sides of the eqs. (\ref{3.17})
and find that for all parameters $P = \{ Z_g, \lambda \}$ we have

\beq
\label{3.18}
{k^2 \over P} \ {dP \over dk^2} < 10^{-10} \quad . 
\eeq       

\noi The same relation holds for all the other parameters $P = \{ Z, n,
m, h, \beta \}$ for $k < k_c$~; this explains, why there is no visible
evolution in this regime. \par 

Strictly speaking, we have not yet obtained an analytic infrared fixed
point of the exact renormalization group equations: the evolutions would
vanish identically for $k \to 0$ only for $\bar{g} = 0$. However, with
$\bar{g} \sim 10^{-12}$ for $k < k_c$ we are so close to this fixed
point that the tiny deviation plays no role for all practically
purposes. (Still, the coupled system of exact renormalization group
equations (\ref{3.13}), (\ref{3.14}) and (3.16) is so complicated that
we did not manage to prove that analytically $\bar{g}(k) \to 0$ for $k
\to 0$). \par 

It may be helpful to get some feeling for the critical scale $k_c$ where
the sudden changes in the evolutions occur. In eq. (\ref{2.28}) in
section 2 we have introduced a scale $M = m/ \sqrt{h}$, which describes
the exponential decay of the correlators of two field strengths $F_{\mu
\nu}$ and thus has some physical meaning. For the ratio $k_c/M$ we
actually obtain, independently of the initial values of $\bar{g}$ and 
$\lambda$, 

\beq
\label{3.19}
k_c/M \sim 10^{-1} \quad .
\eeq

Let us now interpret our results in terms of the effective action
$\Gamma (A, H)$ in the form of eq. (\ref{2.9}). We have to keep in mind
that this truncation constitutes both a low energy approximation (in the
sense that higher derivatives or higher powers of the momenta have been
neglected) and a weak field approximation, since higher powers of
$F_{\mu \nu}$ and $H_{\mu \nu}$ have been omitted.  \par 

First, the fact that the reduced gauge coupling $\bar{g}$ practically
vanishes turns the action into a free quadratic action of $N$ abelian
gauge fields $A_{\mu}^a$ and $N$ antisymmetric tensor fields $H_{\mu
\nu}^a$. This justifies a posteriori the abelian projection, which is
required in order to make the duality transformation (\ref{2.18})
feasible, and the omission of the Schwinger strings during the
calculation of the correlator of two field strengths in
eq.~(\ref{2.21}). \par 

Second, we have introduced two physical dimensionful parameters in
section 2: $\Lambda_c = m/\sqrt{\beta}$ characterizing the $1/p^4$
behaviour of the gluon propagator, and $M = m/\sqrt{h}$ characterizing
the exponential decay of the correlator of two field strengths. Although
the final results of the parameters $m$, $h$, and $\beta$ for $k < k_c$
depend strongly on the initial value of the gauge coupling $g_0$, we
always find $h \sim \beta$ and hence $\Lambda_c \sim M$, and a
dependence of $\Lambda_c$ and $M$ on the bare gauge coupling $g_0$ and 
the starting point $\Lambda$ (which plays implicitly the role of an
ultraviolet cutoff) of the form 

\beq
\label{3.20}
\Lambda_c^2 \sim M^2 \sim \Lambda^2 \ e^{-(16 \pi^2/11g_0^2)} \left ( 1
+ {\cal O}(g_0^2) \right ) \eeq

\noi as it should be. \par

Next we consider the possible definition of a physical gauge coupling in
eq. (\ref{2.13}), $g_{phys} = \bar{g} / \sqrt{Z_{eff}}$ (describing,
e.g., the low energy behaviour of a correlator of two static colored
sources due to dressed one gluon exchange, cf. the second of
refs. \cite{13r}). Both $\bar{g}$ and $Z_{eff}$ are tiny for $k < k_c$,
but to our surprise we find that also the ratio coresponding to
$g_{phys}$ is very small ($\sim 10^{-4}$). This does not imply that Yang
Mills theories are free theories, but rather that a momentum dependent
gauge coupling $\alpha (q^2)$ tends to zero for $q^2 \to 0$ (remember
that we have neglected higher powers of derivatives in our
action). Actually a similar result has been recently found on the
lattice \cite{19r}, and it coincides again with the descriptions of
confinement based on effective abelian models as the dual Higgs model in
\cite{4r}, the ``confining string theory'' based on compact QED (with
monopoles) in \cite{5r}, and the gaussian approximation to field
strength correlators in \cite{6r}.

\mysection{Discussion}
\hspace*{\parindent}
In this paper we have made progress in the description of Yang Mills
theories at low energies in terms of an effective action, which contains
both gluons and an auxiliary field for the field strength as local
fields. It has been known before that, for $Z_{eff} \to 0$, this action
describes a $1/p^4$ behaviour of the gluon propagator and, at the same
time, its abelian projection allows for an explicit duality
transformation relating it to a dual abelian Higgs model describing the
condensation of monopoles. \par 

Here we have shown that it also allows for the computation of the
correlator of two field strengths; the result (\ref{2.27}), however, has
already been obtained before on the basis of the dual Higgs model
\cite{8r} and the confining string theory \cite{9r}. \par

Furthermore we have reconsidered the computation of the corresponding
effective action within the Wilsonian exact renormalization group
approach. We have included a higher dimensional operator in the ghost
sector, which is responsible for the running of the reduced gauge
coupling $\bar{g}$ via the Slavnov-Taylor identities. (The considered 
operator is actually the only one to this order in $F_{\mu \nu}$
consistent with the Slavnov-Taylor identities, if one requires that the
BRST variation of the gluon field is still a total covariant
derivative.) We find that now the Landau singularity in the running gauge
coupling disappears; on the contrary, the running gauge coupling becomes
tiny at small scales. Clearly, the sudden jump of the running coupling
at the critical scale $k_c$ resembles a phase transition; it is not
quite clear, however, whether the same phenomenon would also appear as a
function of the temperature. \par 

Many properties of the resulting quasi-abelian effective action (always
at low momenta, i.e. to lowest order in a derivative expansion) have
already been discussed at the end of section 3. Let us add a final
point, which is due to the smallness of $Z_{eff} = Z - n^2/m^2$: for
$Z_{eff} \to 0$ the first three terms $\sim F^2$, $F \cdot H$ and $H^2$
in  $\Gamma (A, H)$ of eq. (\ref{2.9}) become a perfect square $(Z/4)(F
- mH/\sqrt{Z})^2$. Then, due to the vanishing of $\bar{g}$, a field
redefinition of the form $H \to H' = H + \sqrt{Z}F/m$ makes $\Gamma (A,
H')$ nearly independent of $F$ or the gauge fields $A$: the term $\sim
h$ involving $\partial_{\mu} \widetilde{H}_{\mu \nu}$ (for $\bar{g} =
0$) is invariant under this redefinition of $H$ due to the Bianchi 
identity, only in the term $\sim
\beta$ expressions of the form $\partial_{\mu} F_{\mu \nu}$
appear. Apart from this term $\sim \beta$ the resulting action $\Gamma
(H')$ has then the simple form  

\beq
\label{4.1}
\Gamma (H') = {m^2 \over 4} \left ( H_{\mu \nu} \right )^2 + {h \over 2}
\left ( \partial_{\mu} \ \widetilde{H}_{\mu \nu} \right )^2 
\eeq  

\noi which coincides with the weak field limit of the action of the
universal ``confining string theory'' in \cite{5r}, where the
antisymmetric tensor field couples to the surface bounded by the Wilson
loop like the field strength. \par 

Clearly the relevance of the term $\sim \beta$ has to be better
understood; leaving it aside, a coherent picture emerges: in \cite{5r}
the action (\ref{4.1}) is claimed to be universal in the sense that it
does not depend on details (as the gauge group) of the confining gauge
theory under consideration. Here we have obtained it from the
integration of the Wilsonian exact renormalization group equations in
the infrared regime, which should also be described by universality
classes. \par 

However, within the present approach it is straightforward to include
higher derivative terms or momentum dependent couplings in the effective
action (as, e.g., in the second of refs. \cite{13r}); this allows to
recover the perturbative behaviour of the Green functions at large
momenta. The approach is thus quite unique in allowing for a coherent
description of the effective action both in the perturbative and
non-perturbative regimes. If our results, in particular the infrared
limit in the form of the action (\ref{4.1}), can be confirmed by
including higher dimensional operators in the effective action, a
quantitative treatment of the non-perturbative regime within an
expansion in powers of derivatives for a given Yang Mills theory seems
to be within reach.  

\vspace {4 truecm}
\noi
{\large{\bf Figure Captions}} \par 
\vskip 5 truemm
\begin{description}
\item{\bf Figure 1 :}  Diagrams contributing to the renormalization
group flow $\partial_k Z_g$. Straight lines correspond to ghost
propagators, twiggled lines to gluon propagators, and the crossed circle
denotes an insertion of $\partial_k \widetilde{R}$ with $\widetilde{R}$
as in eq. (\ref{3.12}).\par \vskip 3 truemm 

\item{\bf Figure 2 :} Diagrams contributing to the renormalization group
flow $\partial_k\lambda$.  \par \vskip 3 truemm

\item{\bf Figure 3 :} $log (Z_{eff})$ versus $t = - log (k^2
/\Lambda^2)$. \par \vskip 3 truemm 

\item{\bf Figure 4 :} $\ell n (\bar{g})$ versus $t = - log
(k^2/\Lambda^2)$. \par \vskip 3 truemm

\item{\bf Figure 5 :} $\ell n (\lambda )$ versus $t = - log
(k^2/\Lambda^2)$. \par 
\end{description}

\newpage
\newpage
\def\labelenumi{[\arabic{enumi}]}
\noindent
{\large{\bf References}}
\ben
\item\label{1r} K. G. Wilson, Phys. Rev. {\bf D10} (1974) 2445. 
\item\label{2r} S. Mandelstam, Phys. Rev. {\bf D20} (1979) 3223.   
\item\label{3r} A. M. Polyakov, JETP Lett. {\bf 20} (1974) 194; \\
G. t'Hooft, Nucl. Phys. {\bf B79} (1974) 276.  
\item\label{4r} M. Baker, J. S. Ball and F. Zachariasen, Phys. Rev. {\bf
D37} (1988) 1036 and 3785; Phys. Rev. {\bf 209} (1991) 73; \\
S. Maedan and T. Suzuki, Progr. Theor. Phys. {\bf 81} (1989) 229.   
\item\label{5r} A. M. Polyakov, Nucl. Phys. {\bf B486} (1997) 23; \\
M. C. Diamantici, F. Quevedo and C. A. Trugenberger, Phys. Lett. {\bf
B396} (1997) 115.  
\item\label{6r}H. G. Dosch, Phys. Lett. {\bf B190} (1987) 177; \\
H. G. Dosch and Y. Simonov, Phys. Lett. {\bf B205} (1988) 339; \\
Y. Simonov, Nucl. Phys. {\bf B324} (1989) 67, and hep-ph/9709344.  
\item\label{7r} G. West, Phys. Lett. {\bf B115} (1982) 468.   
\item\label{8r} M. Baker, N. Brambilla, H. G. Dosch and A. Vairo,
HD-THEP-97-60, hep-ph/9802273.     
\item\label{9r} D. V. Antonov, hep-th/9707245 and hep-th/9710144. 
\item\label{10r} U. Ellwanger, LPTHE Orsay 97-44, hep-ph/9710326. 
\item\label{11r}A. Di Giacomo and H. Panagopoulos, Phys. Lett. {\bf
B285} (1992) 133; \\
M. D'Elia, A. Di Giacomo and E. Meggiolaro, Phys. Lett. {\bf B408}
(1997) 315; \\ 
A. Di Giacomo, E. Meggiolaro and H. Panagopoulos, hep-lat/9603017; \\
G. Bali, N. Brambilla and A. Vairo, Phys. Lett. {\bf B421} (1998) 265. 
\item\label{12r} M. Reuter and C. Wetterich, Nucl. Phys. {\bf B417}
(1994) 181 and Phys. Rev. {\bf D56} (1997) 7893; \\
B. Bergerhoff and C. Wetterich, Phys. Rev. {\bf D57} (1998) 1591.
\item\label{13r} U. Ellwanger, M. Hirsch and A. Weber, Z. Phys. {\bf
C69} (1996) 687 and Eur. Phys. J. {\bf C1} (1998) 563.   
\item\label{14r} K. G. Wilson and I. Kogut, Phys. Rep. {\bf 12} (1974)
15; \\ 
F. Wegner, in: Phase Transitions and Critical Phenomena, Vol. 6,
eds. C. Domb and M. Green (Academic Press, NY 1975); \\
J. Polchinski, Nucl. Phys. {\bf B231} (1984) 269; \\
C. Wetterich, Phys. Lett. {\bf B301} (1993) 90; \\
T. Morris, Int. J. Mod. Phys. {\bf A9} (1994) 2411.
\item\label{15r} G. Keller and C. Kopper, Phys. Lett. {\bf B273} (1991)
323; \\ 
C. Becchi, in: Elementary Particles, Field Theory and Statistical
Mechanics, eds. M. Bonini, G. Marchesini and E. Onofri, Parma University
1993; \\ 
M. Bonini, M. D'Attanasio and G. Marchesini, Nucl. Phys. {\bf B418}
(1994) 81, ibid. {\bf B421} (1994) 429, {\bf B437} (1995) 163,
Phys. Lett. {\bf B346} (1995) 87. 
\item\label{16r} U. Ellwanger, Phys. Lett. {\bf B335} (1994) 364. 
\item\label{17r} M. D'Attanasio and T. Morris, Phys. Lett. {\bf B378}
(1996) 213~; \\ 
M. Bonini and F. Vian, Nucl. Phys. {\bf B511} (1998) 479. 
\item\label{18r} U. Ellwanger, C. Wetterich, Nucl. Phys. {\bf B423}
(1994) 137. 
\item\label{19r} Ph. Boucaud et al., Lattice calculation of $\alpha_s$
in momentum scheme, Orsay preprint LPTHE 98/49
 
\een 

\begin{figure}[p]
\unitlength1cm
\begin{picture}(12,20)
\put(-3,-4){\epsffile{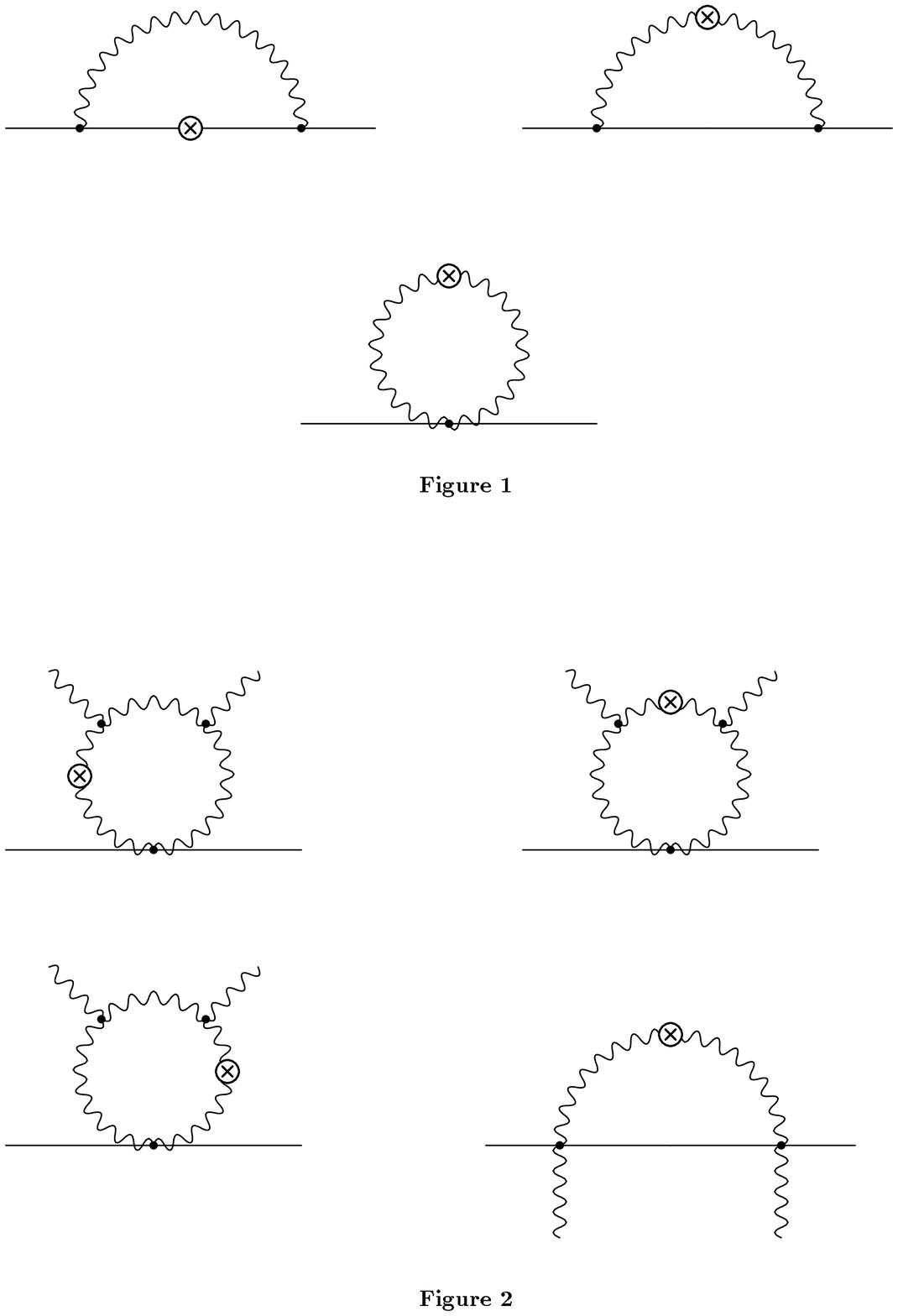}}
\end{picture}
\end{figure}

\begin{figure}[p]
\unitlength1cm
\begin{picture}(12,20)
\put(-3,-5){\epsffile{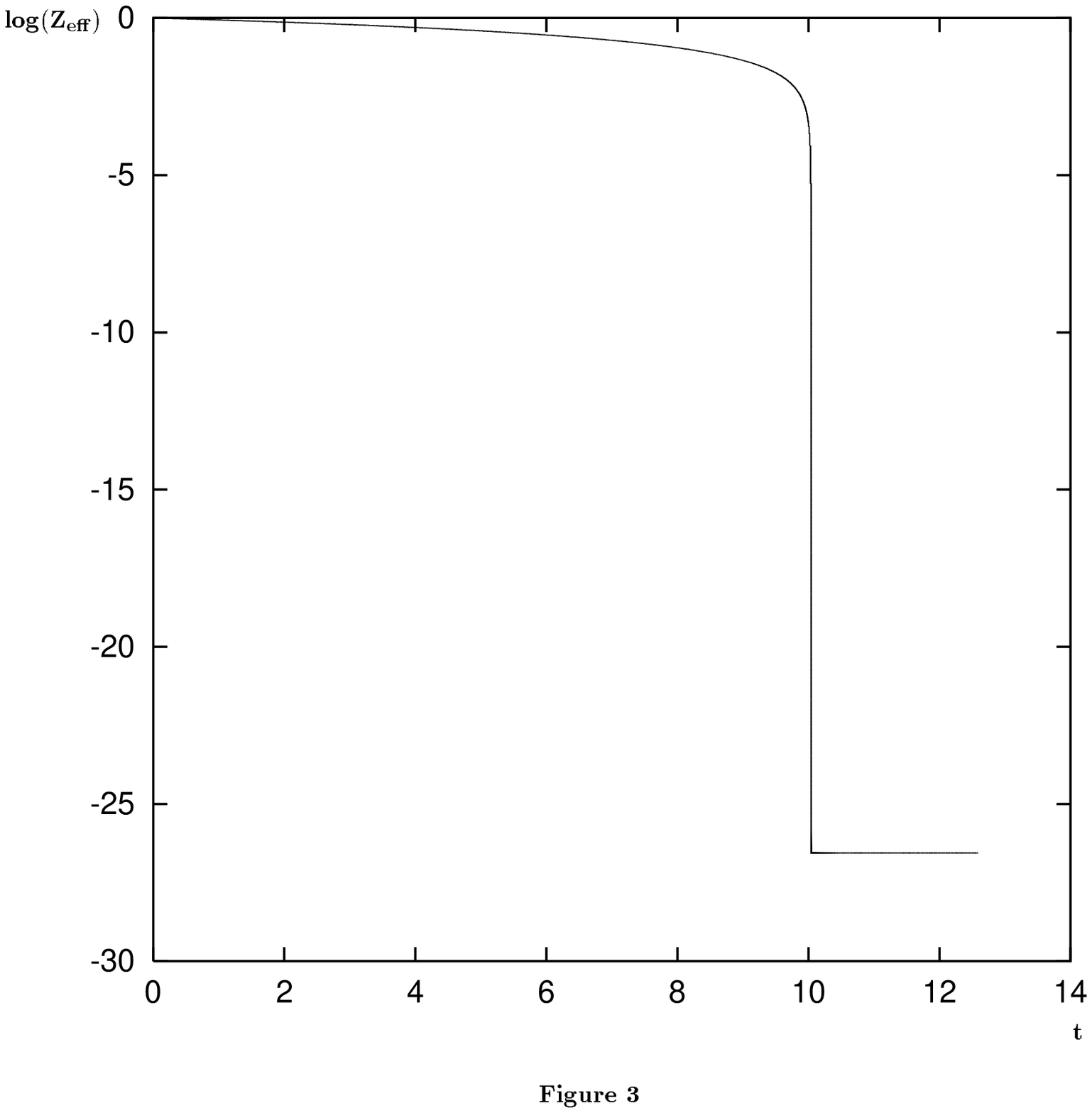}}
\end{picture}
\end{figure}

\begin{figure}[p]
\unitlength1cm
\begin{picture}(12,20)
\put(-3,-5){\epsffile{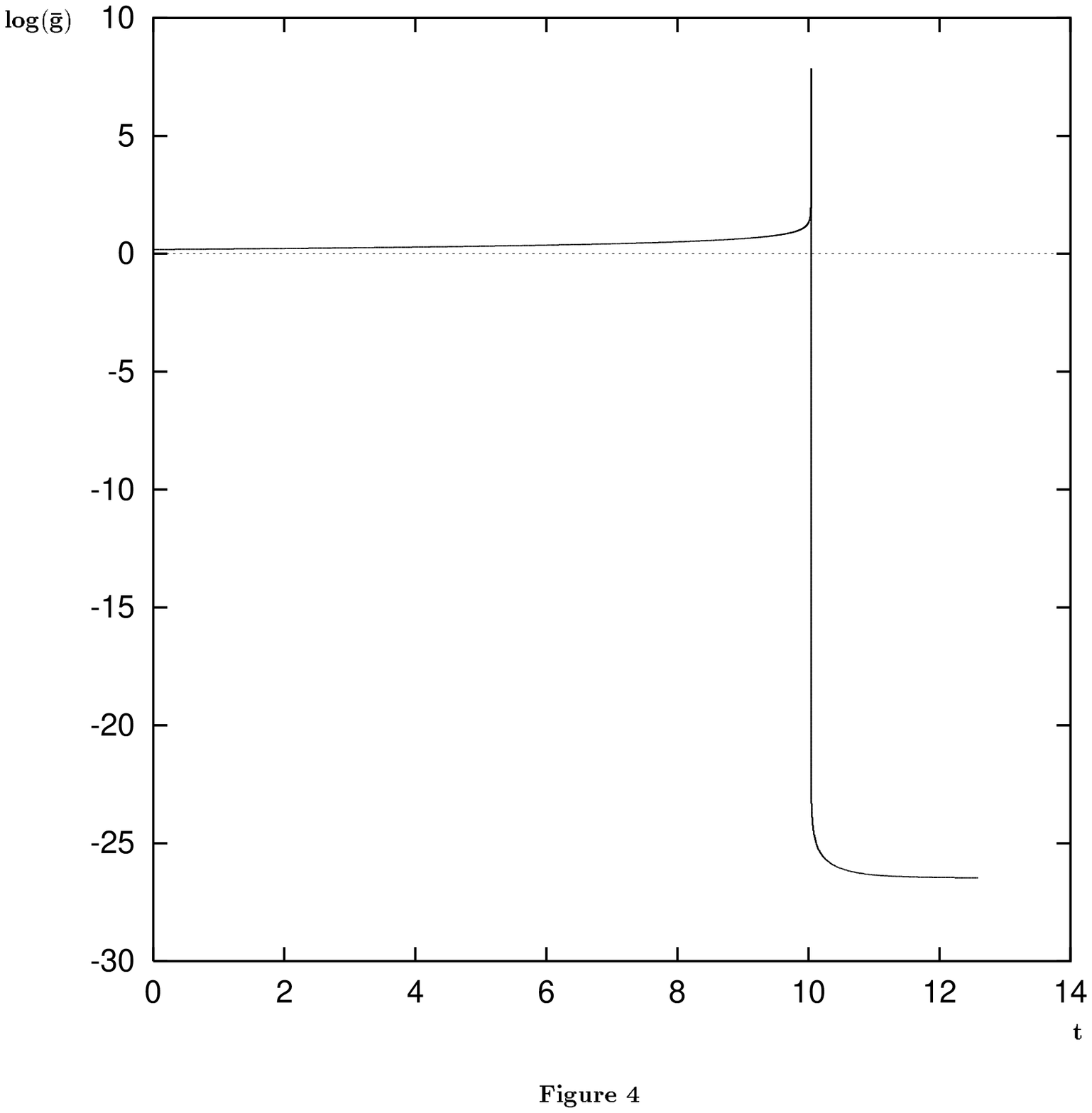}}
\end{picture}
\end{figure}

\begin{figure}[p]
\unitlength1cm
\begin{picture}(12,20)
\put(-3,-5){\epsffile{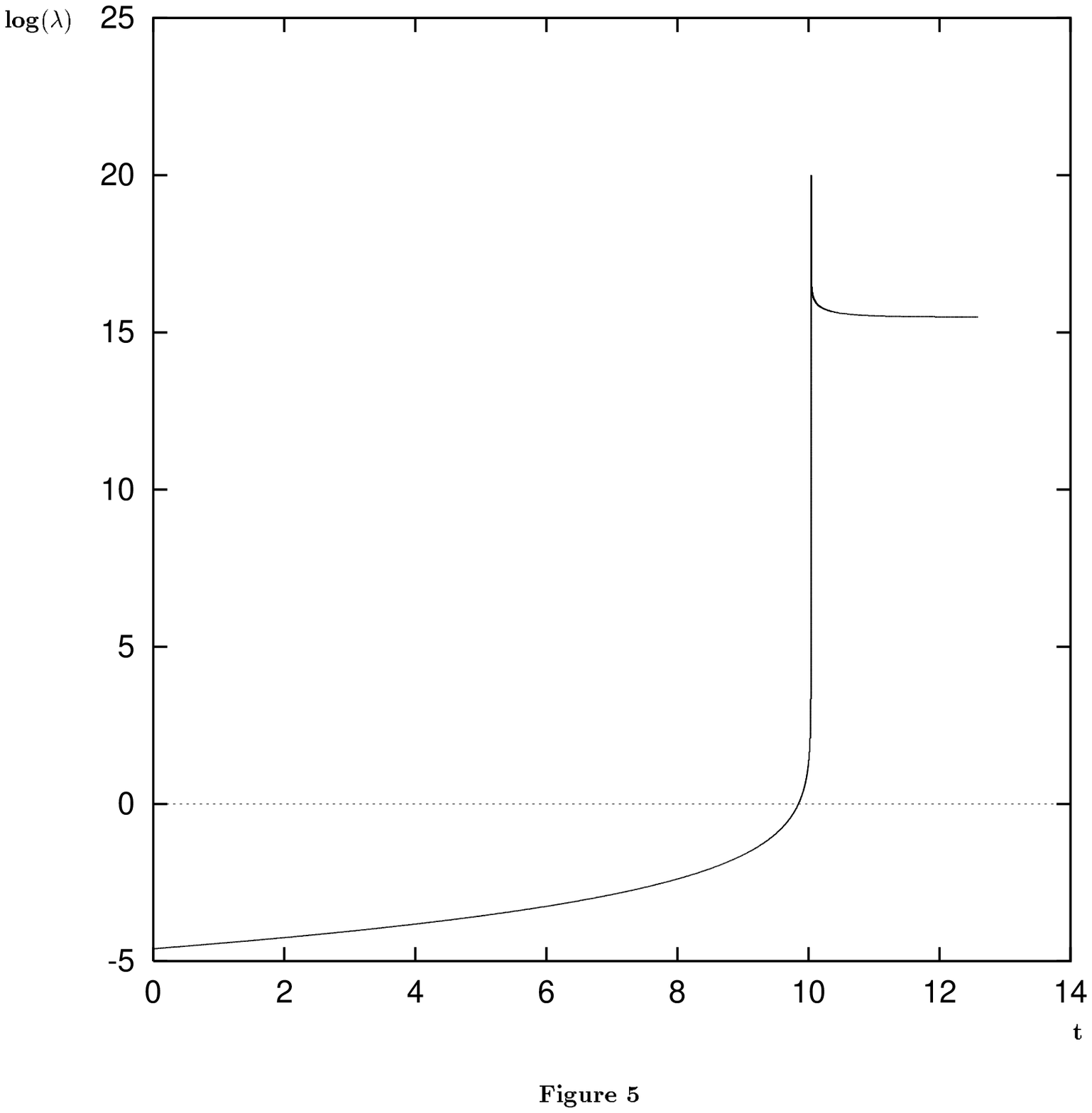}}
\end{picture}
\end{figure}

\end{document}